\documentclass[12pt]{article}
\usepackage{amssymb}
\usepackage{amsmath}
\usepackage{epsfig}
\usepackage{color}

\newcommand{\N}{{\mathcal N}}

\newcommand{\D}{{\mathcal D}}

\renewcommand{\Im}{Im}

\renewcommand{\P}{{\mathsf P}}

\renewcommand{\H}{{\mathcal H}}

\newcommand{\ls}{\lambda(s)}

\renewcommand{\tilde}[1]{\widetilde{#1}}

\newcommand{\bfl}{{\mathbf l}}

\begin{document}

\title{Adiabatic theorems for quantum resonances}
\author{Walid K. Abou Salem\footnote{Current address: Department of Mathematics, Univesity of Toronto, M5S 2E4 Toronto, Canada}
 and J\"urg Fr\"ohlich\\
Institute for Theoretical Physics\\ETH Zurich\\
CH-8093 Zurich, Switzerland\footnote{E-mail: walid@itp.phys.ethz.ch, juerg@itp.phys.ethz.ch}}
\date{ }

\maketitle

\begin{abstract}
We study the adiabatic time evolution of quantum resonances over time scales which are small compared to the lifetime of the resonances. We consider three 
typical examples of resonances: The first one is that of shape resonances corresponding, for example, to the state of a 
quantum-mechanical particle in a potential well whose shape changes over time scales small compared to the escape time of the particle from the well. Our approach to studying the adiabatic 
evolution of shape resonances is based on a precise form of the time-energy uncertainty relation and the 
usual adiabatic theorem in quantum mechanics. The second example concerns resonances that appear 
as isolated complex eigenvalues of spectrally deformed Hamiltonians, such as those encountered in the N-body Stark effect. Our approach to 
study such resonances is based on the Balslev-Combes theory of dilatation-analytic Hamiltonians and an adiabatic 
theorem for nonnormal generators of time evolution. Our third example concerns resonances arising from eigenvalues embedded in the continuous spectrum when a perturbation is turned on, such as those encountered when a small system is coupled to an infinitely extended, dispersive medium. Our approach to this class of examples is based on an extension of adiabatic theorems without a spectral gap condition. We finally comment on resonance crossings, which can be studied using the last approach.
\end{abstract}




\section{Introduction}

There are many physically interesting examples of quantum resonances in atomic physics and quantum optics. To mention one, the state 
of a cold gas of atoms localized in a trap may be metastable, since the trap may be not strictly confining. In 
typical Bose-Einstein condensation experiments, the shape of the trap usually varies slowly over time scales small compared to the lifetime of the metastable state, yet larger than a typical relaxation time (see for example \cite{DGPS}). 
This is an example of an adiabatic evolution of shape resonances. While there has been much progress in 
a time-independent theory of quantum resonances (see \cite{Simon1,Simon2,Hun,Orth,Herbst,CatHunGraf}), there has been relatively little work on a time-dependent theory of quantum resonances (see \cite{FroehPfeif,SofWein,MerSig,Ne}). Surprisingly, and inspite of its relevance to the interpretation of many
experiments and phenomena in atomic physics, the problem of adiabatic evolution of quantum resonances received very litte attention, so far; (but see \cite{Da}). 

In this paper, we study the adiabatic evolution of three general types of quantum resonances. This is a first step towards a rigorous understanding of resonance- and metastability phenomena, such as hysteresis 
in magnets and Sisyphus cooling of atomic gases; (see for example \cite{CA,CT,Ph}). We first consider the adiabatic evolution of so called 
shape resonances. More specifically, we consider a quantum-mechanical particle in a potential well, say that of a quantum dot or a 
locally harmonic trap, with the property that the shape of the potential well changes over time scales which
are small compared to the time needed for the particle to escape from the well. The analysis of this problem is based on a precise
form of the time-energy uncertainty relation, see \cite{FroehPfeif}, and the standard adiabatic theorem in quantum mechanics, \cite{Ka2}. In our approach, we obtain an explicit estimate on the 
distance between the true state of the system and an {\it instantaneous} metastable state. Our approach can also be applied to study the time evolution of the state of an electron in a $He^+$ ion moving in a time-dependent magnetic field which changes over 
time scales that are small compared to the ionization time of the ion; (see \cite{FroehPfeif} for a discussion of this example in the time-independent situation). 

The second class of examples concerns quantum resonances that appear as isolated complex eigenvalues of spectrally deformed Hamiltonians, 
such as the N-body Stark effect; (see for example \cite{Hun,Herbst}).\footnote{For the sake of simplicity, we consider nondegenerate resonances. However, our analysis can be extended to the case of degenerate resonances; (see \cite{MerSig,Ne} for a discussion of the latter in the time-independent case).} Our analysis is based on Balslev-Combes 
theory for dilatation analytic Hamiltonians, \cite{BalComb}, and on an adiabatic theorem for generators of evolution 
that are not necessarily normal or bounded, \cite{AS}. This approach, too, yields explicit estimates on the distance 
between the true state of the system and an instantaneous metastable state. 

The third class of examples concerns resonances that emerge from eigenvalues of an unperturbed Hamiltonian embedded in the continuous spectrum after a perturbation has been added to the Hamiltonian. Typical examples of such resonances arise when a small quantum-mechanical system, say an impurity spin, is coupled to an infinite, dispersive medium, such as magnons (see for example \cite{BFS1,BFS2,BFS3} for relevant physical models). Our approach to such examples is based on an extension of adiabatic theorems {\it without a spectral gap condition}, \cite{Teu,Teu2,AvronElgart,AvronElgart2, ASF}. Our results also cover the case of resonance crossings. 
Further details of applications where our assumptions are explicitly verified for various physical models will appear in \cite{ASF2}.

\bigskip

\noindent{\bf Acknowledgements}

WAS is grateful to an anonymous referee for pointing out references \cite{Da, HJ, J, Ne2}. 


\section{Adiabatic evolution of shape resonances}

In this section, we study the time evolution of the state of a quantum-mechanical particle moving in ${\mathbf R}^d$ 
under the influence of a potential well which is not strictly confining. The potential well is described by a time-dependent function on ${\mathbf R}^d$
\begin{equation}
v_\theta^\tau (x,t)\equiv \theta^2 v(\frac{x}{\theta}, s) ,
\end{equation}
where $\tau$ is the adiabatic time scale, $t$ is the time, $s=\frac{t}{\tau}$ is rescaled time, $\theta\ge 1$ is a parameter characterizing the width and height of the well, and $v(x,s)$ is a function on ${\mathbf R}^d\times{\mathbf R}$ that is twice differentiable in $s\in {\mathbf R}$ and smooth in $x\in {\mathbf R}^d$; see below for precise assumptions on the potential. We assume that $\tau$ is small compared to the escape time of the particle from the well.\footnote{The escape time of the particle from the well, which is related to $\theta,$ will be estimated later in this section.} 
By introducing an auxiliary adiabatic evolution, we obtain precise estimates on the difference between the true
state of the particle and an instantaneous metastable state. Our analysis is based on the generalized time-energy uncertainty relation, as derived in \cite{FroehPfeif}, and on the usual adiabatic theorem in quantum mechanics, \cite{Ka2}. 

The Hilbert space of the system is $\H:=L^2({\mathbf R}^d, d^dx).$ Its dynamics is generated by the time-dependent Hamiltonian
\begin{equation}
\label{Hamiltonian}
H^\tau (t):= -\Delta/2 + v_\theta^\tau (x,t) ,
\end{equation}
where $\Delta$ is the $d$-dimensional Laplacian.\footnote{We work in units where the mass of the particle $m=1,$ and Planck's constant $\hbar=1.$} We make the following assumptions on the potential $v_\theta (x,s),$ for $s\in I,$ where $I$ is an arbitrary, but fixed  compact interval of ${\mathbf R}.$ 
\begin{itemize}
\item[(A1)] The origin $x=0$ is a local minimum of $v(x,s),$ for all $s\in I,$ and, without loss of generality,  $v(0,s)=0$ for $s\in I.$
\item[(A2)] The Hessian of $v(\cdot ,s)$ at $x=0$ is positive-definite, with eigenvalues $\Omega_i^2(s)>\Omega_0^2,i=1,\cdots ,d,$ and $\Omega_0>0$ is a constant independent of $s.$
\item[(A3)] Consider a smooth function $g(x)$ with the properties that $g(x)=1$ for $|x|<\frac{1}{2}$ and $g(x)=0$ for $|x|>1,$ where $|x|:= \sqrt{\sum_{i=1}^d x_i^2}.$ For $\epsilon >0,$ we define the rescaled function $g_{\epsilon ,\theta}$ by 
\begin{equation}
\label{SmoothG}
g_{\epsilon ,\theta}(x):= g(\frac{x}{(\epsilon\theta)^{1/3}}).
\end{equation}
We assume that, for all $\epsilon >0,$
\begin{equation}
\label{A3}
\max_{x\in {\mathbf R}^d} g_{\epsilon, \theta}(x)|v_\theta (x,s)-\frac{1}{2} x\Omega^2(s)x|\le c\epsilon ,
\end{equation}
uniformly in $s\in I$, where $\Omega^2(s)$ is the Hessian of $v(\cdot, s)$ at $x=0$ and $c$ is a finite constant independent of $s\in I.$
\item[(A4)] $v(x,s)$ is smooth, polynomially bounded in $x\in {\mathbf R}^d,$ and bounded from below, uniformly in $s\in I.$ Moreover, $v(x,s)$ is twice differentiable in $s\in I.$ We also assume that $\| H(s_1)-H(s_2)\| \le C, \; \forall s_1,s_2\in I,$ where $C$ is a finite constant.
\end{itemize} 
\noindent Note that under these assumptions, $v_\theta$ is a potential well of diameter of order $O(\theta)$ and height $O(\theta^2).$ Let 
\begin{equation}
\label{H0}
H_0(s):= - \Delta/2 +\frac{1}{2} x \Omega^2(s) x ,
\end{equation}
and
\begin{equation}
\label{H1}
H_1(s):= H_0(s)+w_{\epsilon, \theta}(x,s) , 
\end{equation}
where \footnote{$H_1(s)$ depends on the parameters $\theta$ and $\epsilon,$ but we drop the explicit dependence to simplify notation.}
\begin{equation}
\label{omega}
w_{\epsilon, \theta}(x,s):= g_{\epsilon, \theta} (x)[v_\theta (x,s)-\frac{1}{2}x \Omega^2(s)x ] . 
\end{equation}
Note that 
\begin{equation}
\label{H}
H(s)=H_1(s)+\delta v_{\epsilon,\theta}(x,s) ,
\end{equation}
where 
\begin{equation}
\label{DeltaV}
\delta v_{\epsilon,\theta}(x,s) := (1-g_{\epsilon,\theta}(x))[v_\theta (x,s) -\frac{1}{2}x\Omega^2(s) x ].
\end{equation}
It follows from assumptions (A3) and (A4), and (\ref{DeltaV}) that 
\begin{equation}
\label{SuppV}
\max_{s\in I}|\delta v_{\epsilon,\theta} (x,s) | \le 
\begin{cases}
0, |x|\le \frac{1}{2} (\epsilon\theta)^{1/3} \\
\theta^2 P(x/\theta) , |x| \ge \frac{1}{2}(\epsilon\theta)^{1/3}
\end{cases} ,
\end{equation}
uniformly in $s\in I,$ for some polynomial $P(x)$ of $x.$

Denote by $P_1^n(s),n\in {\mathbf N},$ the projection onto the eigenstates of $H_1(s)$ corresponding to the $n^{th}$ eigenvalue of $H_1(s).$ It follows from assumptions (A3) and (A4) that, for $\epsilon$ small enough, $P_1^n(s)$ is twice differentiable in $s$ as a bounded operator for $s\in [0,1].$

Denote by $U^\tau (s,s')$ the propagator generated by $H(s),$ which solves the equation\footnote{Assumptions (A1)-(A4) are sufficient to show that $U^\tau$ exists as a unique unitary operator with domain $\D,$ a common dense core of $H(s), s\in I.$}  
\begin{equation}
\label{DynamicsEquation}
\partial_s U^\tau (s,s') = -i \tau H(s)U^\tau (s,s'), \; U^\tau(s,s)=1 .
\end{equation}

Suppose that the initial state of the system is given by a density matrix $\rho_0,$ which is a positive trace-class operator with unit trace. Then the state of the particle at time $t=\tau s$ is given by the density matrix $\rho_s,$ which satisfies the Liouville equation 
\begin{equation}
\label{Liouville}
\dot{\rho}_s = -i\tau [H(s), \rho_s]
\end{equation}
and $\rho_{s=0}=\rho_0.$ The solution of (\ref{Liouville}) is given by 
\begin{equation}
\rho_s=U^\tau(s,0)\rho_0U^\tau(0,s) .
\end{equation}

Let $\P$ be an orthogonal projection onto a reference subspace $\P\H,$ and let $p_s$ denote the probability of finding the state of the particle in the reference subspace $\P\H$ at time $t=\tau s.$ This probability is given by
\begin{equation}
\label{Prob1}
p_s:= Tr (\rho_s \P). 
\end{equation}

We are interested in studying the adiabatic evolution of a state of a particle which initially, at time $t=0,$ is localized inside the well. Such a state may be approximated by a superposition of eigenstates of $H_1(0)$ (defined in (\ref{H1})). The initial state of the particle is chosen to be given by 
\begin{equation}
\label{InitialState}
\rho_0 = \sum_{n=0}^N c_n P_1^n (0) ,
\end{equation}
where $P_1^n(0)$ are the eigenprojections onto the states corresponding to the eigenvalues $E_n$ of $H_1(0),$ $c_n\ge 0,$ with $\sum_{n=1}^N c_n =1,$ for some finite integer $N.$ 


We let $U_1$ be the propagator of the auxiliary evolution generated by $H_1(s).$ It is given as the solution of the equation   
\begin{equation}
\partial_s U_1(s,s')=-i\tau H_1(s)U_1(s,s'),\;  U_1(s,s)=1.\label{AuxEvol1}
\end{equation}
Moreover, let
\begin{equation}
\label{WDef}
W(s,0):= U_1(0,s)U^\tau (s,0).
\end{equation}
Then $W(s,0)$ solves the equation
\begin{equation}
\label{AuxiliaryDynamics}
\partial_s W(s,0)=-i\tau \tilde{H}(s)W(s,0), W(0,0)=1.
\end{equation}
where 
\begin{equation}
\label{TildeH}
\tilde{H}(s)= U_1^*(s,0)\delta v_{\epsilon,\theta} (s)U_1 (s,0)
\end{equation}
as follows from (\ref{DynamicsEquation}), (\ref{H1}), (\ref{AuxEvol1}) and (\ref{WDef}). 


Then 
\begin{equation}
p_s = Tr (\rho_s \P)=\sum_n c_n p_s^n, \label{Prob2} 
\end{equation}
where 
\begin{equation}
p_s^n:= Tr (\P U^\tau(s,0)P_1^n(0)U^\tau(0,s)) .\label{pn}
\end{equation}
We define 
\begin{equation}
\label{TildeP}
\tilde{P}_1^n(s) := U_1 (s,0)P_1^n(0) U_1(0,s) .
\end{equation}
We have the following proposition.

\bigskip
\noindent{\bf Proposition 2.1}

{\it Suppose assumptions (A1)-(A4) hold. Then
\begin{equation}
\label{AuxiliaryUncertainty}
p^n_s 
\begin{matrix}
\le \\ \ge 
\end{matrix}
\sin_*^2 (arcsin \sqrt{Tr(\P \tilde{P}_1^n(s))} \pm 2\tau \int_0^s ds' f(P^n_1(0), \tilde{H}(s')) ,
\end{equation}
for $s\ge 0,$ where $p_s^n$ is defined in (\ref{pn}), $\tilde{H}(s)$ in (\ref{TildeH}), $\tilde{P}_1^n(s)$ in (\ref{TildeP}),
\begin{equation}
sin_*(x):=
\begin{cases}
0 , x<0 \\
sin(x), 0\le x\le \frac{\pi}{2} \\
1, x>\frac{\pi}{2}
\end{cases} ,
\end{equation}
and 
\begin{equation}
f(P,A):= \sqrt{Tr (PA^*(1-P)A)}.
\end{equation}}
\bigskip

The proof of Proposition 2.1 is given in the Appendix, and it is based on the generalized time-energy uncertainty relation derived in \cite{FroehPfeif}.

Before stating an adiabatic theorem for shape resonances, we want to estimate the time needed for the quantum-mechanical particle to escape from the potential well if its initial state is given by (\ref{InitialState}). Note that, for each {\it fixed} value of $s\in I,$ the spectrum of $H_0(s),$ $\sigma(H_0(s)),$ is formed of the eigenvalues 
\begin{equation}
\label{SpecH0}
E^s_{\bfl}=\sum_{i=1}^d\Omega_i (s) (l_i + \frac{1}{2}),
\end{equation}
where $\bfl = (l_1,\cdots,l_d)\in {\mathbf N}^d,$ with corresponding eigenfunctions 
\begin{equation}
\label{Hermite}
\phi_\bfl ^s (x)= \prod_{i=1}^d \Omega_i^{1/4}(s)h_{l_i}(\sqrt{\Omega_i(s)}x_i) ,
\end{equation}
where $h_{l_i}$ are Hermite functions normalized such that $$\int dx h_l(x)h_k(x)=\delta_{lk}.$$ Recall that the Hermite functions decay like a Gaussian away from the origin,
\begin{equation}
\label{HermiteDecay}
|h_l(x)|\le c_{l,\delta} e^{-(\frac{1}{2}-\delta)x^2} ,
\end{equation}
for an arbitrary $\delta>0$ and a finite constant $c_{l,\delta};$ (see for example \cite{RS}). It follows from analytic perturbation theory (Lemma A.1 in the Appendix) that the eigenstates of $H_1(s)$ decay like a Gaussian away from the origin. Moreover, it follows from  assumption (A3) that $\delta v_{\epsilon,\theta}$ is supported outside a ball of radius $\frac{1}{2}(\epsilon\theta)^{1/3}.$ 

Let $\pi^n_1(x,y;s)$ denote the kernel of $U_1(s,0)P_1^n(0)U_1(0,s),$ whose modulus decays like a Gaussian away from the origin for arbitrary finite $\tau$; see Lemma A.1 in the Appendix. For each fixed $s\in I,$ the following estimate follows from Lemma A.1, assumptions (A3)-(A4) and (\ref{TildeH}).
\begin{eqnarray}
f(P_1^n(0),\tilde{H}(s))^2 &=&
|Tr (P_1^n(0)\tilde{H}(s)^2-P_1^n(0)\tilde{H}(s)P_1^n(0)\tilde{H}(s))|\nonumber \\ &=& |Tr([P_1^n(0),\tilde{H}(s)]^2)|\nonumber \\
&=& |Tr([U_1(s,0)P_1^n(0)U_1(0,s), \delta v_{\epsilon,\theta}(s)]^2)|\nonumber\\
&=& \int dx dy |\pi^n_1(x,y;s)|^2 (\delta v_{\epsilon,\theta}(x,s)- \delta v_{\epsilon,\theta}(y,s))^2 \nonumber \\
&\le& C_{\epsilon,n } e^{-\mu_\epsilon \theta^{2/3}},\label{Lifetime}
\end{eqnarray}
where $\mu_\epsilon$ is proportional to $\epsilon^{2/3},$ $C_{\epsilon , n}$ is a finite constant independent of $s\in I,$ (for finite $n$ appearing in (\ref{InitialState}) and fixed $\epsilon$). Let 
\begin{equation}
\label{LifeTime1}
\tau_l\sim e^{\mu_\epsilon\theta^{2/3}/2},
\end{equation}
which, by (\ref{Lifetime}) and (\ref{InitialState}), is a lower bound for the time needed for the particle to escape from the well. \footnote{In other words, the particle spends an exponentially large time in $\theta$ inside the well. Note that one may also directly use time-dependent perturbation theory to estimate the time needed for the particle to escape from the well, see \cite{FroehPfeif}.}

We now introduce the generator of the adiabatic time evolution for each eigenprojection,
\begin{equation}
\label{HAdiab}
H_a^n(s):= H_1(s)+\frac{i}{\tau}[\dot{P}_1^n(s),P_1^n(s)] ,
\end{equation} 
and the corresponding propagator $U_a^n(s,s')$ which satisfies 
\begin{equation}
\label{UadiabEq}
\partial_s U_a^n(s,s')=-i\tau H_a^n (s)U_a^n(s,s'); \; U_a^n(s,s)=1 .
\end{equation} 
By assumptions (A1)-(A4), it follows that (\ref{UadiabEq}) has a unique solution, $U_a^n(s,s'),$ which is a unitary operator. From the standard adiabatic theorem in quantum mechanics \cite{Ka2}, we know that 
\begin{eqnarray}
U_a^n (s,s') P_1^n (s') U_a^n(s',s) &=& P_1^n(s) \label{IntertwiningProp} ,\\
\sup_{s\in [0,1]}\| U_a^n(s,0)-U_1(s,0)\| &=& O(\tau^{-1}),\label{AdiabTheorem1}
\end{eqnarray} 
for $\tau\gg 1.$\footnote{We work in units where a microscopic relaxation time is of order unity.}

For $1\ll \tau\ll\tau_l,$ where $\tau_l$ is given in (\ref{LifeTime1}), it follows from (\ref{AuxiliaryUncertainty}), (\ref{TildeP}), (\ref{IntertwiningProp}) and (\ref{AdiabTheorem1}) that 
\begin{equation}
p_s^n = Tr(\P P_1^n(s)) + O(\max(1/\tau,\tau/\tau_l)).\label{AuxiliaryUncertainty2}
\end{equation}

Let 
\begin{equation}
\tilde{\rho}_s := \sum_{n=1}^N c_n P_1^n(s),\label{InstMeta}
\end{equation}
the {\it instantaneous} metastable state of the particle inside the well. 

By (\ref{InitialState}), (\ref{AuxiliaryUncertainty2}) and (\ref{InstMeta}), we have that, for
\begin{equation}
1\ll \tau\ll\tau_l ,
\label{TauChoice}
\end{equation} 
\begin{equation}
\sup_{s\in [0,1]}|p_s - Tr (\P \tilde{\rho}_s)|\le  A /\tau + B\tau/\tau_l ,
\end{equation}
where $A$ and $B$ are finite constants. This proves the following theorem for the adiabatic evolution of shape resonances.

\bigskip 

\noindent{\bf Theorem 2.2 (Adiabatic evolution of shape resonances)} 

{\it Suppose assumptions (A1)-(A4) hold for some $\tau$ satisfying (\ref{TauChoice}). Then 
\begin{equation}
p_s = Tr (\P\tilde{\rho}(s)) + O(\max(\frac{1}{\tau},\frac{\tau}{\tau_l})) .
\end{equation}}

In other words, over time scales that are small compared to the escape time $\tau_l,$ given in (\ref{LifeTime1}), of the particle from the potential well, the true state of the particle which is initially localized inside the well, as given by the choice (\ref{InitialState}), is approximately equal to the instantaneous metastable state given in (\ref{InstMeta}). We remark that a similar analysis can be applied to study the adiabatic evolution of the metastable state of the electron of an $He^+$ ion moving in a time-dependent magnetic field; (see \cite{FroehPfeif} for a discussion of this model in the time-independent case).


\section{Isolated eigenvalues of spectrally deformed Hamiltonians}

In this section, we discuss the adiabatic evolution of quantum resonances which appear as isolated eigenvalues of spectrally deformed Hamiltonians. Examples of such resonances include ones of the Stark effect and the N-body Stark effect (see for example \cite{Hun,Herbst,Simon1,Simon2,Da}). Our analysis is based on Balslev-Combes theory for dilatation analytic Hamiltonians and on an adiabatic theorem for nonnormal and unbounded generators of evolution. The main result of this section is Theorem 3.3, which gives an estimate on the distance between the true state and an instantaneous metastable state when the adiabatic time scale is much smaller than the lifetime of the metastable state.

\subsection{Approximate metastable states}

Consider a quantum mechanical system with Hilbert space $\H$ and a family of selfadjoint Hamiltonians $\{ H^\tau_g(t)\}_{t\in {\mathbf R}},$ which are given by 
\begin{equation}
H_g^\tau(t)=H_g(s),
\end{equation}
with fixed dense domain of definition, where
\begin{equation}
H_g(s)=H_0(s) + g V(s) ,
\end{equation}
and $H_0(s)$ is the (generally time-dependent) unperturbed Hamiltonian, while $gV (s)$ is a perturbation bounded relative to $H_0(s),$ unless specified otherwise; see footnote after assumption (B1) below. Here, $s=t/\tau\in [0,1]$ is the rescaled time. Let $U(\theta), \theta \in {\mathbf R},$ denote the one-parameter unitary group of dilatations. For fixed $g,$ we assume that there exists a positive $\beta ,$ independent of $s\in [0,1],$ such that  
\begin{equation}
\label{DeformedHamiltonian}
H_g(s,\theta ):= U(\theta) H_g(s)U(-\theta) ,
\end{equation}
extends from real values of $\theta$ to an analytic family in a strip $|\Im\theta|<\beta,$ for all $s\in [0,1].$ The spectrum of $H_g(s, \theta)$ is assumed to lie in the closed lower half-plane for $\Im\theta \in (0,\beta).$ The relation 
\begin{equation}
H_g(s,\theta)^* = H_g(s,\overline{\theta}) 
\end{equation}
holds for real $\theta$ and extends by analyticity to the strip $|\Im\theta|<\beta .$ We make the following assumptions.

\begin{itemize}

\item[(B1)] $\lambda_0(s)$ is an isolated or embedded simple
 eigenvalue of $H_0(s)$ with eigenprojection $P_0(s).$\footnote{The Stark effect for discrete eigenvalues of Coulumb systems is an example where isolated eigenvalues of the unperturbed Hamiltonian become resonances once the unbounded perturbation is turned on, \cite{Hun2, HerbstSimon}.} We assume that, for each fixed $s\in [0,1]$ and $\Im\theta\in (0,\beta),$ $\lambda_0(s)$ is separated from the essential spectrum of $H_0(s,\theta ).$ We also assume that the corresponding eigenprojection $P_0(s,\theta)$ is analytic in $\theta$ for $\Im\theta\in (0,\beta)$ and strongly continuous in $\theta$ for $\Im\theta\in [0,\beta).$ 

\begin{center}
\begin{picture}(0,0)%
\includegraphics{fig1.pstex}%
\end{picture}%
\setlength{\unitlength}{2072sp}%
\begingroup\makeatletter\ifx\SetFigFont\undefined%
\gdef\SetFigFont#1#2#3#4#5{%
  \reset@font\fontsize{#1}{#2pt}%
  \fontfamily{#3}\fontseries{#4}\fontshape{#5}%
  \selectfont}%
\fi\endgroup%
\begin{picture}(5289,3361)(3589,-6029)
\put(7966,-3526){\makebox(0,0)[lb]{\smash{{\SetFigFont{6}{7.2}{\familydefault}{\mddefault}{\updefault}{\color[rgb]{0,0,0}$2\Im\theta$}%
}}}}
\put(5806,-2851){\makebox(0,0)[lb]{\smash{{\SetFigFont{6}{7.2}{\familydefault}{\mddefault}{\updefault}{\color[rgb]{0,0,0}$\lambda_0(s)$}%
}}}}
\put(5941,-3751){\makebox(0,0)[lb]{\smash{{\SetFigFont{6}{7.2}{\familydefault}{\mddefault}{\updefault}{\color[rgb]{0,0,0}$\lambda_g(s)$}%
}}}}
\end{picture}%

\end{center}

\item[(B2)] For $0<\Im \theta<\beta,$ let $\tilde{H}_g(s,\theta) = P_g(s,\theta)H_g(s,\theta)P_g(s,\theta)$ denote the reduced Hamiltonian acting on $Ran (P_g(s,\theta)),$ and let $\lambda_g(s)$ be its corresponding eigenvalue. Then 
$$\lambda_g(s)\stackrel{g\rightarrow 0}{\longrightarrow}\lambda_0(s).$$ We assume that $\lambda_g(s)$ is differentiable in $s\in [0,1].$

\item[(B3)] For each fixed $s\in [0,1]$ and fixed $\theta$ with $\Im\theta\in (0,\beta),$ there is an annulus ${\mathcal N}(s,\theta)\subset{\mathbf C}$ centered at $\lambda_0(s)$ such that the resolvent,
\begin{equation}
R_g(s,\theta;z) := (z-H_g(s,\theta))^{-1},
\end{equation}
exists for each $z\in {\mathcal N}(s,\theta)$ and $0\le g < g_0(z).$

\item[(B4)] Let $\gamma(s)$ be an arbitrary contour in ${\mathcal N}(s,\theta)$ enclosing $\lambda_0(s)$ and $\lambda_g(s),$ for $\Im\theta\in (0,\beta).$ Then, for $0\le g<g(\gamma(s)),$ the spectral projection
\begin{equation}
\label{DefProj}
P_g(s,\theta):= \oint_{\gamma(s)} \frac{dz}{2\pi i} R_g(s,\theta ;z)
\end{equation}
satisfies
\begin{equation}
\label{ProjDiff2}
\lim_{g\rightarrow 0} \| P_g(s,\theta) - P_0(s,\theta) \| = 0.
\end{equation}
We assume that $P_g(s,\theta)$ is twice differentiable in $s\in [0,1]$ as a bounded operator, for fixed $\theta, \Im\theta\in (0,\beta).$

\item[(B5)] {\it RS (Rayleigh-Schr\"odinger) Expansion.} The perturbation $V(s,\theta),$ for $|\Im\theta|<\beta,$ is densely  defined and closed, and $V(s,\theta)^*=V(s,\overline{\theta}).$ 
We define 
$$H_g(s,\theta):=H_0(s,\theta)+gV(s,\theta)$$ 
on a core of $H_g(s,\theta).$ For $\Im\theta\ne 0, z\in {\mathcal N}(s,\theta)$ and $g$ small enough, the iterated resolvent equation is
\begin{equation}
\label{ResolventEq}
R_g(s,\theta ;z)P_0(s,\theta) = \sum_{n=0}^{N-1} g^n R_0(s,\theta;z) A_n(s,\theta;z) +g^N R_g(s,\theta;z)A_N(s,\theta;z) ,
\end{equation}
for $N\ge 1$ (depending on the model), where 
\begin{equation}
\label{ResolventTerms}
A_n(s,\theta;z):= (V(s,\theta)R_0(s,\theta;z))^n P_0(s,\theta ).
\end{equation}
We assume that the individual terms in (\ref{ResolventEq}) are well-defined, and that $A_n(s,\theta;z)$ defined in (\ref{ResolventTerms}) are analytic in $\theta$ in the strip $\Im\theta\in (0,\beta),$ for $n=1,\cdots ,N,$ and $z\in {\mathcal N}(s,\theta),$ and strongly continuous in $\Im\theta\in[0,\beta).$
 This assumption is satisfied for $N=1$ in dilatation-analytic systems where $V(s,\theta)$ is bounded relative to $H_0(s,\theta),\Im\theta\in [0,\beta)$; see, e.g., \cite{Simon1,Simon2}. Moreover, this assumption holds for arbitrary $N\ge 1,$ if $\lambda_0(s)$ is an isolated eigenvalue of the unperturbed Hamiltonian $H_0(s),$ as in the case of discrete eigenvalues of Coulumb systems, with $V(s,\theta)$ a perturbation describing the Stark effect, \cite{Hun2, HerbstSimon}. 

\end{itemize}

The RS-expansion for $P_g(s,\theta)$ implies that, for $\Im\theta\in (0,\beta),$ 
\begin{equation}
\label{ApproxProj}
P_g(s,\theta)=P_g^N (s,\theta)+O(g^N),
\end{equation}
where $P_g^N(s,\theta)$ is analytic in the strip $\Im\theta\in (0,\beta),$ and strongly continuous in $\Im\theta\in[0,\beta).$

In other words, the spectral projection onto the resonance state is only defined up to a certain order $N$ in the coupling constant $g.$ This is to be expected since resonance states decay with time. We now show that, for each {\it fixed} $s\in [0,1],$ the projections $P_g^N(s)$ can be regarded as projections onto approximate metastable states, up to an error of order $O(g^N)$. 

Denote by $\psi_0(s)$ the eigenstate of $H_0(s)$ with corresponding eigenvector $\lambda_0(s),$ and let 
\begin{equation}
\label{PsiN}
\psi_g^N(s)= \frac{1}{\|P_g^N(s)\psi_0(s)\|} P_g^N(s) \psi_0(s).
\end{equation}
We have the following proposition for approximate metastable states, for each {\it fixed} $s\in [0,1];$ see \cite{Hun}.

\bigskip
\noindent {\bf Proposition 3.1 (Approximate metastable states)}

{\it Assume that (B1)-(B5) hold, and fix $s\in [0,1].$ Let $\xi\in C_0^\infty ({\mathbf R})$ be supported close to $\lambda_0(s)$ with $\xi=1$ in some open interval containing $\lambda_0(s).$ Then 
\begin{equation}
\langle \psi_g^N(s), e^{-iH_g(s) t} \xi (H_g(s))  \psi_g^N(s) \rangle = a_g^N(s)e^{-i\lambda_g(s) t} + b_g^N(t) ,\label{DecayIsolated}
\end{equation}
for small $g,$ where 
\begin{equation*}
a_g^N(s)=\langle \psi_g^N(s,\overline{\theta}), P_g(s,\theta)\psi_g^N(s,\theta)\rangle = 1+O(g^{2N}), \Im\theta\in (0,\beta),
\end{equation*}
and
\begin{equation*}
b_g^N(t)\le g^{2N} C_m (1+t)^{-m}, 
\end{equation*}
for $m> 0,$ where $C_m$ is a finite constant, independent of $s\in [0,1].$
}
\bigskip

Although the proof of Proposition 3.1 is a straightforward extension of the results in \cite{Hun}, it is sketched in the Appendix to make the presentation self-contained.

Choosing $t=0$ in (\ref{DecayIsolated}) gives 
\begin{equation}
\langle \psi_g^N(s) , (1 - \xi(H_g(s)))\psi_g^N(s)\rangle = O(g^{2N}).
\end{equation}
In particular, for $0<\xi\le 1,$
\begin{equation}
\langle \psi_g^N(s), e^{-iH_g(s) t}\psi_g^N(s)\rangle = e^{-i\lambda_g(s) t}+ O(g^{2N}).
\end{equation}
This motivates considering $\psi_g^N(s)$ as approximate {\it instantaneous} metastable states, up to an error term of order $O(g^{2N}).$ 

In the next subsection, we recall a general adiabatic theorem proven in \cite{AS}.


\subsection{A general adiabatic theorem}

Consider a family of closed operators $\{ A (t) \}_{t\in {\mathbf
R}}$ acting on a Hilbert space $\H,$ with common dense domain of definition $\D.$ Let $U(t)$ be the propagator given by
\begin{equation}
\partial_t U(t) \psi= -A(t)U(t) \psi\; , \; U(t=0) =1 \; ,
\label{timeevolution}
\end{equation}
for $t\ge 0 ; \; \psi\in\D.$ We make the following assumptions, which will be verified in the application we consider later in this section.

\begin{itemize}
\item[(C1)] $U(t)$ is a bounded semigroup, for $t\in {\mathbf R}^+,$ ie, $\|U(t)\|\le M,$ where $M$ is a finite constant.

\item[(C2)] For $z\in\rho(A(t)),$ the resolvent set of $A(t)$, let $R(z,t):= (z-A(t))^{-1}$. Assume that $R(-1,t)$ is bounded and differentiable as a bounded operator on $\H$, and that $A(t)\dot{R}(-1,t)$ is bounded,
where the $(\dot{} )$ stands for differentiation with respect to
$t$.

\end{itemize}

Assume that $A(t)\equiv A(0)$ for $t\le 0$, and that it is
perturbed {\it slowly} over a time scale $\tau$ such that
$A^{(\tau)}(t)\equiv A(s)$, where $s:=\frac{t}{\tau}\in [0,1] $ is
the rescaled time. The following two assumptions are needed to
prove an adiabatic theorem.

\begin{itemize}

\item[(C3)] The eigenvalue $\ls\in \sigma(A(s))$ is isolated and
simple, with
$$dist(\ls, \sigma(A(s))\backslash \{ \ls\})>\delta,$$
where $\delta>0$ is a constant independent of $s\in [0,1]$, and
$\ls$ is continuously differentiable in $s\in [0,1]$.

\item[(C4)] The projection onto $\ls$,
\begin{equation}
P_\ls:= \frac{1}{2\pi i}\oint_{\gamma_\ls} R(z,s) dz \; ,
\end{equation}
where $\gamma_\ls$ is a contour enclosing $\ls$ only, is twice
differentiable as a bounded operator.

\end{itemize}

Note that, since $\ls$ is simple, the resolvent of $A(s)$ in a neighborhood $\N$ of $\ls ,$
contained in a ball ${\mathbf B}(\ls,r)$ centered at $\ls$ with
radius $r<\delta ,$ is
\begin{equation}
\label{res} R(z,s)=\frac{P_\ls}{z-\ls} + R_{analytic} (z,s) \; ,
\end{equation}
where $R_{analytic}(z,s)$ is analytic in $\N$.

We now discuss our general adiabatic theorem. Let $U_\tau(s,s')$ be the propagator given by
\begin{equation}
\label{rtevol2}
\partial_s U_\tau (s,s') = -\tau A(s) U_\tau (s,s')\; , U_\tau(s,s)=1 \; ,
\end{equation}
for $s\ge s'.$
Moreover, define the generator of the {\it adiabatic time
evolution},
\begin{equation}
A_a(s):= A(s)-\frac{1}{\tau}[\dot{P}_\ls,P_\ls] \; ,
\end{equation}
with the corresponding propagator $U_a(s,s'),$ which is given by
\begin{equation}
\label{ate2}
\partial_s U_a(s,s') = - \tau A_a(s) U_a (s,s') \; ; U_a(s,s)=1 \; ,
\end{equation}
for $s\ge s'.$
It follows from assumption (C4) that 
\begin{equation*}
\sup_{s\in[0,1]}\|[\dot{P}_\ls, P_\ls]\|\le C,
\end{equation*}
for some finite constant $C$, and hence by perturbation theory for semigroups, \cite{Ka1} chapter IX, and assumption (C1), 
$U_a$ defined on the domain $\D$ exists and is unique, and $\|U_a(s,s')\| < M'$ for $s\ge s'$, where $M'=Me^{C}.$ We are in a position to state our adiabatic theorem.

\bigskip

\noindent {\bf Theorem 3.2 (A general adiabatic theorem)}

{\it Assume (C1)-(C4). Then the following holds.

\begin{itemize}
\item[(i)]
\begin{equation}
\label{intertwining}P_\ls U_a (s,0) =  U_a (s,0)P_\lambda (0) \;
,
\end{equation}
for $s\ge 0$ (the intertwining property).
\item[(ii)] $$\sup_{s\in [0,1]} \| U_\tau (s,0)-U_a(s,0) \| \le \frac{C}{1+\tau},$$ for $\tau>0$ and $C$ a finite constant. In particular, 
$$\sup_{s\in [0,1]} \| U_\tau (s,0)-U_a(s,0) \|=O(\tau^{-1}),$$
for $\tau\gg 1.$

\end{itemize}}

We refer the reader to \cite{AS} for a proof of Theorem 3.2.

\bigskip
\noindent{\bf Remark.} Assumption (C1) can be relaxed, but the result of Theorem 3.2 will be weakened. Suppose $A(t)$ generates a quasi-bounded semigroup, ie, there exist finite positive constants $M$ and $\gamma$ such that $\|U(t)\| \le Me^{\gamma t}, t\in {\mathbf R}^+,$  then (ii) in Theorem 3.2 becomes
\begin{equation*}
\sup_{s\in [0,1]}\| U_\tau (s,0)- U_a(s,0)\| \le C \frac{e^{\tau\gamma}}{\tau},
\end{equation*}
for $1\ll \tau\ll \gamma^{-1}.$

\subsection{Adiabatic evolution of resonances that appear as isolated eigenvalues of spectrally deformed Hamiltonians}

We consider a quantum mechanical system satisfying assumptions (B1)-(B5), subsection 3.1. Denote by $U_\tau(s,s',\theta)$ the propagator corresponding to the deformed time evolution, which is given by 
\begin{equation}
\label{DefEvol1}
\partial_s U_\tau (s,s',\theta) = -i\tau H_g(s,\theta) U_\tau (s,s',\theta), U_\tau (s,s,\theta)=1,
\end{equation}
for $0\le s'\le s\le 1$ and $\Im\theta\in [0,\beta).$ We make the following assumption on the existence of the deformed time evolution, which can be shown to hold in specific physical models; see \cite{ASF2, Hun, Hun2} and \cite{Ka1}, chapter IX.

\begin{itemize}
\item[(B6)] For fixed $\theta$ with $\Im\theta\in(0,\beta),$ $U_\tau(s,s',\theta), 0\le s'\le s\le 1,$ exists and is unique as a bounded semigroup with some dense domain of definition $\D.$\footnote{We remark later how this assumption can be relaxed.} In particular, there exists a finite constant $M$ such that
$$\| U_\tau(s,s',\theta)\| \le M, 0\le s'\le s\le 1.$$  
\end{itemize}
The generator of the deformed adiabatic time evolution is given by
\begin{equation}
H_a(s,\theta) := H_g(s,\theta) + \frac{i}{\tau} [\dot{P}_g(s,\theta),P_g(s,\theta)],
\end{equation}
and it generates the propagator 
\begin{equation}
\label{DefEvol2}
\partial_s U_a (s,s',\theta) = -i\tau H_a(s,\theta)U_a(s,s',\theta) , \; U_a(s,s,\theta)=1,
\end{equation}
for $0\le s'\le s\le 1$ and fixed $\theta$ with $\Im\theta\in(0,\beta).$

For fixed $\theta$ with $\Im\theta\in(0,\beta),$ assumptions (B4) and (B6) and perturbation theory for semigroups, \cite{Ka1}, imply that $U_a(s,s',\theta),s\ge s',$ exists and 
\begin{eqnarray}
\\
\| U_a(s,s',\theta)\| &\le& M' ,
\end{eqnarray}
where $M'$ is a finite constant independent of $s,s'\in [0,1].$

Assumptions (B1)-(B6) in subsection 3.1 imply assumptions (C1)-(C4) in subsection 3.2, with the identification 
\begin{align*}
H_g(s,\theta)& \leftrightarrow\ -iA(s)\\
\lambda_g(s)& \leftrightarrow -i\lambda (s),\\
P_g(s,\theta)&\leftrightarrow iP_\ls ,\\
\end{align*}
for fixed $\theta$ with $\Im\theta\in(0,\beta).$ 

We consider a reference subspace corresponding to a projection $\P$ which is dilatation analytic, ie, $\P(\theta)=U(\theta)\P U(-\theta)$ extends from real values of $\theta$ to a family in a strip $|\Im\theta|<\beta, \beta>0.$
Moreover, we assume that the initial state of the quantum mechanical system is 
\begin{equation}
\label{InitialState2}
\rho_0= |\psi_g^N(0)\rangle\langle\psi_g^N(0)|,
\end{equation}
where $\psi_g^N(s)$ has been defined in (\ref{PsiN}).

We are interested in estimating the difference between the true state of the system and the {\it instantaneous} metastable state defined in (\ref{PsiN}) when $H_g$ varies over a time scale smaller than the lifetime of the metastable state, 
$$\tau_l=\min_{s\in [0,1]}(\Im \lambda_g(s))^{-1}\sim g^{-2}.$$ 
More precisely, we are interested in comparing 
\begin{eqnarray}
p_{\tau s}&:=& Tr (\P U_\tau(s,0)\rho_0 U^*_\tau(s,0)) \nonumber \\ &=&Tr (\P U_\tau (s,0)|\psi_g^N(0)\rangle\langle\psi_g^N(0)|U_\tau^*(s,0)) 
\end{eqnarray}
to
\begin{equation}
\tilde{p}_{\tau s}:= Tr (\P |\psi_g^N(s)\rangle\langle\psi_g^N(s)|).
\end{equation}
This is given in the following theorem.

\bigskip

\noindent {\bf Theorem 3.3 (Adiabatic evolution of isolated resonances)}

{\it Suppose assumptions (B1)-(B6) hold. Then, for $g$ small enough and for $1\ll \tau\ll \tau_l \sim g^{-2},$
\begin{equation}
|p_{\tau s}-\tilde{p}_{\tau s}| = O(\max(1/\tau, g^N\tau, \tau/\tau_l(g))).
\end{equation}
}

\bigskip

{\it Proof.} This result is a consequence of Theorem 3.2. Since assumptions (B1)-(B6) hold, we know that, for fixed $\theta$ with $\Im\theta\in (0,\beta),$  
\begin{eqnarray}
U_a(s,0,\theta) P^N_g(0,\theta)&=&P^N_g(s,\theta)U_a(s,0,\theta) +O(g^N\tau)\label{Intertwining} \\
\sup_{s\in [0,1]}\| U_a (s,0,\theta)-U_\tau (s,0,\theta)\| &\le& \frac{C}{\tau},\label{PropagatorDiff}
\end{eqnarray}
for $\tau \gg 1,$ where $C$ is a finite constant. For $1\ll \tau\ll g^{-2},$ and $\Im\theta\in (0,\beta),$we have 
\begin{align*}
p_{\tau s} &= \langle U_\tau (s,0) \psi_g^N(0), \P U_\tau (s,0)\psi_g^N(0)\rangle \\
&= \langle U_\tau (s,0,\overline{\theta}) \psi_g^N(0,\overline{\theta}), \P(\theta)U_\tau(s,0,\theta)\psi_g^N(0,\theta)\rangle \\
&= \langle U_a (s,0,\overline{\theta}) \psi_g^N(0,\overline{\theta}), \P(\theta) U_a(s,0,\theta) \psi_g^N(0,\theta)\rangle + O(1/\tau) \\
&= \langle \psi_g^N(s,\overline{\theta}), \P(\theta) \psi_g^N(s,\theta)\rangle + O(\max (\tau/\tau_l(g), 1/\tau, g^N\tau )) \\
&= \tilde{p}_{\tau s} +  O(\max(1/\tau,g^N\tau, \tau/\tau_l(g))).
\end{align*} $\Box$ 

\bigskip
\noindent {\it Remarks.}
\begin{itemize}
\item[(1)] To estimate the survival probability of the true state of the system, choose $\P=|\psi_g^N(0)\rangle\langle \psi_g^N(0)|,$ where $\psi_g^N(s)$ is defined in (\ref{PsiN}).

\item[(2)] One may also estimate the difference between the true expectation value of a bounded operator $A$ and its expectation value in the instantaneous metastable state, provided the operator $A$ is dilatation analytic. Similar to the proof of Theorem 3.3, one can show that 
\begin{equation}
\langle \psi_g^N(0), U_\tau (s,0)^* A U_\tau (s,0)\psi_g^N(0)\rangle = \langle \psi_g^N(s), A \psi_g^N(s)\rangle + O(\max(1/\tau,g^N\tau, \tau/\tau_l(g))),\nonumber
\end{equation}
for $1\ll \tau\ll g^{-2}.$

\item[(3)] The results of this section can be extended to study the quasi-static evolution of equilibrium and nonequilibrium steady states of quantum mechanical systems at positive temperatures, e.g., when one or more thermal reservoirs are coupled to a small system with a finite dimensional Hilbert space; see \cite{AS,ASF} for further details. In these applications, the generator of time evolution is deformed using complex translations instead of complex dilatations.

\item[(4)] Assumption (B6) can be relaxed. Fix $\theta$ with $\Im\theta\in (0,\beta).$ Suppose that 
$H_g(s,\theta)$ generates a quasi-bounded semigroup,
\begin{equation}
\label{QuasiBdSemiGr}
\| U_\tau(s,s',\theta) \| \le M e^{g\alpha\tau (s-s')},
\end{equation}
where $M$ and $\alpha$ are positive constants and $g$ is the coupling constant. It follows from assumption (B4) that
\begin{equation*}
\frac{1}{\tau}\sup_{s\in[0,1]}\| [\dot{P}_g(s,\theta)P_g(s,\theta)]\| \le \frac{C}{\tau}
\end{equation*} 
for finite $C.$ Together with (\ref{QuasiBdSemiGr}), this implies that
\begin{equation*}
\| U_a(s,s',\theta) \| \le M' e^{g\alpha\tau (s-s')},
\end{equation*}
where $M'$ is a finite constant. 
Then, under assumptions (B1)-(B6), the result of Theorem 3.3 becomes
\begin{equation*}
|p_{\tau s}-\tilde{p}_{\tau s}| = O(\max(e^{g\alpha\tau}/\tau, g^N\tau, \tau/\tau_l(g))),
\end{equation*}
for $1\ll\tau\ll g^{-2}.$

\item[(5)] The results of this section can be extended to study {\it ``superadiabatic''} evolution of quantum resonances. \footnote{We are grateful to an anonymous referee for indicating this possibility to us.} In the last decade, there has been a lot of progress in studying superadiabatic processes (see for example \cite{HJ} and references therein). Depending on the smoothness of the generator of the time evolution, superadiabatic theorems give improved estimates of the difference between the true time evolution and the adiabatic one. Very recently, and after the submission of this paper, superadiabatic theorems with a gap condition have been extended to evolutions generated by nonselfadjoint operators \cite{J}. Using superadiabatic theorems and methods developed in \cite{Ne2}, the results of this section can be extended to longer time scales under additional regularity assumptions on the Hamiltonian. Further details will appear in \cite{ASF2}.

\end{itemize}


\section{General  resonances}

In this section, we study the case of resonances which emerge from eigenvalues of an unperturbed Hamiltonian embedded in the continuous spectrum after a perturbation has been added to the Hamiltonian. Such resonances arise, for example, when a small system, say a toy atom or impurity spin, is coupled to a quantized field, e.g. to magnons or the electromagnetic field. The main result of this section is Theorem 4.1, which is based on an extension of the adiabatic theorem without a spectral gap; see for example \cite{Teu,AvronElgart,ASF}. The results of this section are more general than section 3, since the perturbation is not restricted to be dilatation analytic. 

Consider a quantum mechanical system with a Hilbert space $\H$ and a family of time-dependent selfadjoint Hamiltonians $\{ H_g(t)\}_{t\in {\mathbf R}}$ such that 
\begin{equation*}
H_g(t)=H_0(t)+gV(t),
\end{equation*}
where $H_0(t)$ is the unperturbed Hamiltonian with fixed common dense domain of definition $\D, \forall t\in {\mathbf R},$ and $V(t)$ is a perturbation which is bounded relative to $H_0(t)$ in the sense of Kato.\cite{Ka1} We assume that the variation of the true Hamiltonian, $H_g^\tau(t),$ in time is given by $H_g^\tau (t)\equiv H_g(s),$ where $s\in [0,1]$ is the rescaled time. We make the following assumptions on the model.

\begin{itemize}

\item[(D1)] $H_g(s)$ is a generator of a contraction semigroup for $s\in [0,1]$ with fixed dense core. Let $R_g (z,s):= (z-H_g(s))^{-1}$ for $z\in \rho(H_g(s)),$ the resolvent set of $H_g(s).$ We assume that $R_g(i,s)$ is differentiable in $s$ as a bounded operator, and $H_g(s)\dot{R}_g(i,s)$ is bounded uniformly in $s\in [0,1].$ This assumption is sufficient to show that the unitary propagator generated by $H_g(s)$ exists and is unique.

\item[(D2)] $\lambda_0(s)$ is a simple eigenvalue of $H_0(s)$ which is embedded in the continuous spectrum of $H_0(s),$ with corresponding eigenvector $\phi (s),$ 
\begin{equation*}
H_0(s)\phi (s)=\lambda_0(s)\phi(s).
\end{equation*}
Furthermore, the eigenprojection $P_0(s)$ corresponding to $\lambda_0(s)$ is twice differentiable in $s$ as a bounded operator {\it for almost all} $s\in [0,1],$ and is continuous in $s,s\in [0,1],$ as a bounded operator.

\item[(D3)] Let $\overline{P}_0(s):= 1-P_0(s),$ and, for a given operator $A$ on $\H,$ denote by $\hat{A}_s$ its restriction to the range of $\overline{P}_0(s),$
\begin{equation*}
\hat{A}_s := \overline{P}_0(s)A \overline{P}_0(s).
\end{equation*}
Let 
\begin{equation}
\label{LevShift}
F(z,s):= \langle \phi(s), V(s)\overline{P}_0(s)(z-\hat{H}_0(s)_s)^{-1}\overline{P}_0(s)V(s)\phi(s)\rangle .
\end{equation}
For each $s\in [0,1],$ we have
\begin{equation}
\Im F(\lambda_0(s)+i0,s) \le 0, (Fermi's \; Golden \; Rule).
\end{equation}
We note that 
\begin{align*}
P_0(s)H_g(s) &= \lambda_0(s)P_0(s)+O(g)\\
H_g(s)P_0(s) &= \lambda_0(s)P_0(s)+O(g).
\end{align*}

\item[(D4)] {\it Instantaneous metastable states.} Let $\xi\in C_0^\infty({\mathbf R})$ be supported in a neighborhood of $\lambda_0(s).$ For each {\it fixed} $s\in [0,1],$ we have
\begin{equation}
\label{QuasiExpDecay}
\langle \phi (s), e^{-it H_g(s)}\xi(H_g(s))\phi(s)\rangle = a_g(s)e^{-it\lambda_g(s)} + b_g(t) , t\ge 0 ,
\end{equation}
where 
\begin{equation*}
\lambda_g(s)=\lambda_0(s)+ g\langle \phi(s), V(s) \phi (s)\rangle + g^2 F(\lambda_0(s)-i0,s) + o(g^2),
\end{equation*}
and
\begin{eqnarray*}
| a_g(s)-1 | &\le& C g^2 ,\\
| b_g(t) | &\le& 
C g^2 (1+t)^{-n} ,
\end{eqnarray*}
$C$ is a finite constant independent of $s\in [0,1],$ for some $n\ge 1.$ 
Note that $\Im \lambda_g(s)\le 0.$ Equation (\ref{QuasiExpDecay}) uniquely defines the {\it instantaneous} resonance state, up to an error $O(g^4).$ \footnote{The latter assumption is satisfied if the following holds, for each fixed $s\in [0,1]$; see \cite{CatHunGraf} for a proof of this claim in the $s$-independent case.
\begin{itemize}
\item[(1)] There exists a selfadjoint operator $A_s$ such that 
\begin{equation*}
e^{itA_s}\D\subset \D,
\end{equation*}
for each fixed $s\in [0,1]$ and $t\in {\mathbf R}.$ This implies that $\D\cap \D(A_s)$ is a core of $H_0(s).$ 

\item[(2)] Denote by $ad^j_{A_s}(\cdot):=[A_s,ad_{A_s}^{j-1}], ad_{A_s}^1(\cdot) := [A_s,\cdot].$ For some integer $m\ge n+6,$ where $n$ appears in (D4), the multiple commutators $ad_{A_s}^{i}(H_0(s))$ and $ad_{A_s}^i(V(s)),i=1,\cdots, m,$ exist as $H_0(s)$-bounded operators in the sense of Kato. \cite{Ka1}

\item[(3)] {\it Mourre's inequality} holds for some open interval $\Delta_s \ni \lambda_0(s),$
\begin{equation*}
E_{\Delta_s}(H_0(s))i [H_0(s),A_s] E_{\Delta_s}(H_0(s)) \ge \theta E_{\Delta_s }(H_0(s)) + K,
\end{equation*}
where $E_{\Delta_s}(H_0(s))$ is the spectral projection of $H_0(s)$ onto $\Delta_s,$ $\theta$ is a positive constant, and $K$ is a compact operator. 

\end{itemize}}

\end{itemize}

A physical example where assumptions (D1)-(D4) may be satisfied is a small system interacting with a field of noninteracting bosons or fermions, for example, a spin system coupled to a time-dependent magnetic field; see \cite{BFS1,BFS2,BFS3,ASF2} for further details on the relevant model of a toy atom interacting with the electromagnetic radiation field. 

We are interested in the adiabatic evolution of the quantum resonance over time scales which are much smaller than the lifetime of the resonance. We will prove an adiabatic theorem {\it without a spectral gap condition} for quantum resonances for weak coupling $g$ (see \cite{Teu,AvronElgart,AvronElgart2,ASF}).

Let $U_\tau(s,s')$ be the propagator given by
\begin{equation}
\partial_s U_\tau (s,s') = -i\tau H_g(s) U_\tau (s,s'), \; U_\tau (s,s)=1 ,
\end{equation}
with some dense domain of definition $\D.$ Eixstence of $U_\tau$ as a unique unitary operator follows from assumption (D1) and Theorem X.70 in \cite{RS}. Moreover, we introduce the generator of the adiabatic time evolution  
\begin{equation}
H_a^0(s):= H_g(s)+\frac{i}{\tau}[\dot{P}_0(s),P_0(s)]. 
\end{equation}
The propagator corresponding to the approximate adiabatic evolution is given by
\begin{equation}
\partial_s U_a^0(s,s') = -i\tau H_a^0 (s) U^0_a(s,s'), \; U_a^0(s,s)=1,
\end{equation}
with domain of definition $\D.$ Note that $U_a$ exists as a unique unitary operator due to assumptions (D1) and (D2). We have the following theorem, which is an extension of the results in \cite{Teu,AvronElgart,ASF}.

\bigskip

\noindent{\bf Theorem 4.1 (Adiabatic theorem for embedded resonances)} 

{\it Suppose assumptions (D1)-(D4) hold. Then, for small enough coupling $g$ and large enough $\tau,$
\begin{equation}
\label{Intertwining2}
U_a^0(s,0)P_0(0)U_a^0(0,s)= P_0(s) + O(\tau g),
\end{equation}
and 
\begin{equation}
\sup_{s\in [0,1]} \| U_\tau (s,0)- U_a^0(s,0)\| \le \frac{A}{\tau^{1/2}} + Bg\tau^{1/4} + C(\tau^{-1/4}),\label{C2}
\end{equation}
where $A$ and $B$ are finite constants, and $C(x)$ is a positive function of $x\in {\mathbf R}$ such that $\lim_{x\rightarrow 0}C(x)=0.$ In particular, choosing $\tau\sim g^{-2/3}$ gives
\begin{equation}
\sup_{s\in [0,1]}\| U_\tau(s,0)P_0(0)-P_0(s)\| \le A g^{1/3} + C(g^{1/6}).\label{C3}
\end{equation}
}

\bigskip

{\it Proof.} Let
\begin{equation}
h(s,s'):= U_a^0(s,s') P_0(s') U_a^0(s',0).
\end{equation}
Then
\begin{align*}
\partial_{s'} h(s,s') &= i\tau U_a^0(s,s') \{ H_a^0(s')P_0(s')-P_0(s')H_a^0(s')\} U_a^0(s',0) \\
&= i\tau U_a^0(s,s')\{ \lambda_0(s')P_0(s') + \frac{i}{\tau}\dot{P}_0(s')P_0(s') - \lambda_0(s') P_0(s')\\&+ \frac{i}{\tau}P_0(s')\dot{P}_0(s') + O(g)\} U_a^0(s',0) \\
&= O(g\tau) ,
\end{align*}
where we have used the definition of the generator of the adiabatic evolution and the property that 
\begin{equation*}
\dot{P}_0(s)P_0(s)+P_0(s)\dot{P}_0(s)=0.
\end{equation*}
It follows that $h(s,0)=h(s,s),$ which is claim (\ref{Intertwining2}).

Moreover, we are interested in estimating the difference between the true evolution and the adiabatic time evolution. For $\psi\in \D ,$ we have that
\begin{align*}
(U_\tau (s,0)-U_a^0(s,0))\psi &= - \int_0^s ds'\partial_{s'} (U_\tau(s,s')U_a^0(s',0))\psi \\
&= -i\tau \int_0^s ds' U_\tau(s,s')[H_g(s')-H_a^0(s')]U_a^0(s',0)\psi \\
&= -\int_0^s ds' U_\tau (s,s')[\dot{P}_0(s'),P_0(s')]U_a^0(s',0) \psi .
\end{align*}
Since the domain of definition $\D$ is dense in $\H,$ it follows that 
\begin{equation}
\| U_\tau (s,0) - U_a^0(s,0)\| = \| \int_0^s ds' U_\tau(s,s')[\dot{P}_0(s'),P_0(s')]U_a^0(s',0)\| .
\end{equation}

We will now use a variant of Kato's commutator method to express the integrand as a total derivative plus a remainder term; see \cite{Teu}. Let 
\begin{equation}
X_{\epsilon} (s):= R_{g}(\lambda_0 (s)+i\epsilon,s) \dot{P}_0(s)P_0(s)+ P_0(s)\dot{P}_0(s)R_{g}(\lambda_0(s)-i\epsilon,s) .
\end{equation}
Note that 
\begin{align*}
[H_g(s),X_{\epsilon}(s)] &= [H_g(s)-\lambda_0(s)-i\epsilon, R_{g}(\lambda_0(s)+i\epsilon,s)\dot{P}_0(s)P_0(s)] \\ &+ [H_g(s)-\lambda_0(s)+i\epsilon, P_0(s) \dot{P}_0(s)R_{g}(\lambda_0(s)-i\epsilon,s)] \\
&= [\dot{P}_0(s),P_0(s)] +i\epsilon X_\epsilon (s) + O(g/\epsilon).
\end{align*}
Furthermore, 
\begin{align*}
\partial_{s'} (U_\tau (s,s')X_{\epsilon}(s')U_a^0(s',0)) &= i\tau U_\tau (s,s')[H_g(s'),X_{\epsilon}(s')]U_a^0(s',0) \\&+ U_\tau(s,s')X_{\epsilon}(s')[\dot{P}_0(s'),P_0(s')]U_a^0(s',0)\\
&+ U_\tau (s,s')\dot{X}_{\epsilon}(s')U_a^0 (s',0).
\end{align*}
Therefore,
\begin{eqnarray}
\| \int_0^s ds' U_\tau (s,s')[\dot{P}_0(s'),P_0(s')]U_a^0(s',0)\| &\le& \sup_{s\in [0,1]} \{ \frac{1}{\tau}[ \| X_{\epsilon}(s)\| (1+2 \| \dot{P}_0(s)P_0(s)\|)  \nonumber \\
& +& \| \dot{X}_{\epsilon}(s)\| ] + \epsilon \| X_{\epsilon}(s)\| \} + Cg/\epsilon ,
\end{eqnarray}
where $C$ is a finite constant independent of $s\in [0,1].$
We claim that the following estimates are true for small enough $\epsilon$ and $g.$ 
\begin{eqnarray}
&(i)& \| X_{\epsilon}(s)\| < C/\epsilon\\
&(ii)& \| \dot{X}_{\epsilon}(s)\| < C/\epsilon^2\\
&(iii)& \epsilon \| X_{\epsilon}(s)\| < B(\epsilon)+ Cg/\epsilon,\label{Est3} 
\end{eqnarray}
where $\lim_{\epsilon\rightarrow 0}B(\epsilon)=0,$ and $C$ is a finite constant, uniformly in $s\in [0,1].$

Estimates $(i)$ and $(ii)$ follow from our knowledge of the spectrum of $H_g(s)$ and the resolvent identity. To prove estimate $(iii),$ we compare the LHS of (\ref{Est3}) to the case when $g=0.$ Let
\begin{equation}
\tilde{X}_{\epsilon}(s):= R_{0}(\lambda_0(s)+i\epsilon,s)\dot{P}_0(s)P_0(s)+P_0(s)\dot{P}_0(s)R_0(\lambda_0(s)-i\epsilon,s).
\end{equation} 
Then, by the second resolvent identity, 
\begin{equation*}
\|X_{\epsilon}(s)\|\le \| \tilde{X}_{\epsilon}(s)\| + Cg/\epsilon^2 ,
\end{equation*}
uniformly in $s,$ for some finite constant $C.$
We claim that 
\begin{equation}
\lim_{\epsilon\rightarrow 0} \epsilon^2 \| \tilde{X}_{\epsilon}(s)\|^2 = 0.
\label{SmallE}
\end{equation} 
Consider $\phi\in\D,$ then $\psi(s)=\dot{P}_0(s)P_0(s)\phi\in Ker (P_0(s)).$ Using the spectral theorem for $H_0(s),$ we have the following result.
\begin{align*}
\lim_{\epsilon\rightarrow 0} \epsilon^2 \| R_0(\lambda_0(s)+i\epsilon,s)\dot{P}_0(s)P_0(s)\phi\|^2 &= \lim_{\epsilon\rightarrow 0}\epsilon^2 \langle \psi (s), R_{0}(\lambda_0(s)-i\epsilon,s)R_{0}(\lambda_0(s)+i\epsilon,s)\psi(s)\rangle \\
&= \lim_{\epsilon\rightarrow 0}\epsilon^2 \int d\mu_{\psi(s)}(\lambda)\frac{1}{(\lambda-\lambda_0(s))^2+\epsilon^2} \\
&=\mu (\psi(s)\in Ran (P_0(s))) =0,
\end{align*}
and hence claim (\ref{SmallE}).
Therefore,
\begin{equation}
\sup_{s\in [0,1]} \| U_\tau(s,0)-U_a^0(s,0)\| \le \frac{C_1}{\tau \epsilon^2}+\frac{C_2g}{\epsilon}+C(\epsilon),
\end{equation}
where $C_{1,2}$ are finite constants, and $\lim_{\epsilon\rightarrow 0}C(\epsilon)=0.$
Choosing $\epsilon = \tau^{-1/4}$ gives (\ref{C2}). By choosing $\tau\sim g^{-2/3},$ (\ref{C3}) follows from assumption (D4), (\ref{Intertwining2}) and (\ref{C2}).

$\Box$

\bigskip

\noindent{\it Remarks.} 

\begin{itemize}

\item[(1)] We note that, using an argument due to Kato, \cite{Ka2}, the case of finitely many resonance crossings is already covered by Theorem 4.1, since the latter holds for $P_0(s)$ twice differentiable as a bounded operator {\it for almost all} $s\in [0,1]$ and continuous as a bounded operator for $s\in [0,1].$ Suppose that at time $s_0\in [0,1]$, a crossing of $\lambda_0(s)$ with an eigenvalue of $H_0(s)$ happens.  It follows from continuity of $P_0(s)$ that, for small $\epsilon>0,$ $Ran (P_0(s_0-\epsilon))$ and $Ran (P_0(s_0+\epsilon))$ are close up to an error which is arbitrarily small in $\epsilon,$ and hence our claim follows.

\item[(2)] Further knowledge of the spectrum of $H_0(s)$ will yield a better estimate of the convergence of $\epsilon\| \tilde{X}_\epsilon\|$ to zero as $\epsilon\rightarrow 0.$ For example, it is shown in \cite{Teu, Teu2, AvronElgart} that if the spectral measure $\mu_{\phi(s)},\phi(s)\in Ran (P_0(s)),$ is $\alpha$-H\"older continuous, for $\alpha\in [0,1],$ uniformly in $s\in [0,1],$ then \footnote{A measure $\mu$ is $\alpha$-H\"older continuous, $\alpha\in [0,1],$ if there exists a finite constant $C$ such that, for every set $\epsilon$ with Lebesgue measure $|\epsilon|<1,$ $\mu(\epsilon)< C |\epsilon|^{\alpha}$, see, e.g., \cite{Ka1}.} 
\begin{equation}
\sup_{s\in[0,1]}\epsilon \| R_0(\lambda_0(s)+i\epsilon,s)\dot{P}_0(s)P_0(s)\| \le A \epsilon^{\alpha/2},
\end{equation} 
for $\epsilon$ small enough, where $A$ is a finite constant, and hence estimate (\ref{C3}) becomes 
\begin{equation}
\sup_{s\in [0,1]}\| U_\tau(s,0)P_0(0)-P_0(s)\| = O(g^{\alpha/12})
\end{equation}
for $g$ small enough.

\end{itemize}

\section{Appendix}

\noindent{\bf Proof of Proposition 2.1, Section 2}

\noindent {\it Proof of Proposition 2.1.} This proposition effectively follows by integrating the Liouville equation and applying the Cauchy-Schwarz inequality. It is a special case of the generalized time-energy uncertainty relations derived in \cite{FroehPfeif}. Consider an orthogonal projection $P$ and selfadjoint operators $A$ and $B$ acting on a Hilbert space $\H.$ Then it follows from a direct application of the Cauchy-Schwarz inequality that
\begin{equation}
Tr(P[A,B])^2 \le 4 Tr(PA^2 -PAPA) Tr(PB^2-PBPB),\label{Uncertainty1}
\end{equation} 
with equality when there exist $a,b\in {\mathbf R}\backslash \{0\}$ such that
\begin{equation}
[aA+ibB,P]P=0.
\end{equation}
We use inequality (\ref{Uncertainty1}) to derive upper and lower bounds for $p_s^n.$ Let 
\begin{equation}
p^n_{s,s'}:= Tr (\P U_\tau (s,s') P_1^n(0) U_\tau (s',s)). 
\end{equation}
Then 
\begin{align*}
|\partial_{s'}p_{s,s'}^n| &= |i \tau Tr (\P U_\tau (s,s')[H(s'),P_1^n(0)]U_\tau (s',s))|\\
&= |\tau Tr (P_1^n(0)[U_\tau (s',s)\P U_\tau (s,s'),H(s')])|\\
&\le 2 \tau Tr (U_\tau(s,s')P_1^n(0)U_\tau (s',s)\P^2 - \\ &- U_\tau(s,s')P_1^n(0)U_\tau (s',s)\P U_\tau(s,s')P_1^n(0)U_\tau (s',s)\P)^{1/2}\times\\
&\times Tr(P_1^n(0)H(s')^2-P_1^n(0)H(s')P_1^n(0)H(s'))^{1/2} \\
&\le 2 \tau \sqrt{p_{s,s'}^n - (p_{s,s'}^{n})^2} f(P_1^n(0),H(s')),
\end{align*}
where $f(P,A):=\sqrt{Tr(PA^*(1-P)A)}.$
It follows that
\begin{align*}
|\int_0^s ds' \frac{\partial_{s'}p_{s,s'}^n}{\sqrt{p_{s,s'}^n-(p_{s,s'}^n)^2}}| &=|arcsin(\sqrt{p_{s,0}^n})-arcsin(\sqrt{p_{s,s}^n})| \\ 
&\le 2 \tau \int_0^s ds' f(P_1^n(0),H(s')) ,
\end{align*}
and hence 
\begin{equation}
p_s^n= p_{s,0}^n 
\begin{matrix}
\le\\ \ge 
\end{matrix}
sin^2_*(arcsin(\sqrt{Tr(\P P_1^n(0))} \pm 2\tau \int_0^s ds' f(P_1^n(0)),H(s'))) .\label{GenUncertainty1}
\end{equation}
We note that 
\begin{equation}
p_s^n= Tr (\P U_\tau (s,0)P_1^n(0)U_\tau(0,s)) = Tr(U_1(0,s) \P U_1(s,0) W(s,0)P_1^n(0)W(0,s)) .
\end{equation}
Together with (\ref{GenUncertainty1}), and the identification
\begin{align*}
\P&\leftrightarrow U_1(0,s)\P U_1(s,0) \\
H(s)&\leftrightarrow \tilde{H}(s),\\
\end{align*}
where $\tilde{H}(s)$ is the generator of the auxiliary propagator $W$, as defined in (\ref{TildeH}), we have 
\begin{equation}
p_s^n 
\begin{matrix}
\le \\ \ge 
\end{matrix}
sin_*^2 (arcsin \sqrt{Tr (\P U_1(s,0)P_1^n(0)U_1(0,s))}\pm 2\tau \int_0^s ds' f(P_1^n(0),\tilde{H}(s'))).
\end{equation}
$\Box$
\bigskip


\noindent{\bf Proof that the eigenstates of $H_1$, Section 2, decay like a Gaussian in space}

It follows from assumption (A3) that $\| w_{\epsilon,\theta}(x,s)\|$ defined in (\ref{omega}) is uniformly bounded by $c\epsilon ,$ for $s\in I.$ Therefore, the spectrum of $H_1(s),$ for each {\it fixed} $s\in I,$ can be computed by applying analytic perturbation theory(see, e.g., \cite{Ka1,RS}). Also using analytic perturbation theory, one can show that the eigenstates of $H_1(s),$ for each fixed $s,$ decay like a Gaussian away from the origin; (see \cite{CombesThomas,Hun2}). To prove the last claim, choose $E>0.$ 
There exist finitely many sequences $\bfl^{(1)},\cdots , \bfl^{(k_E)},$ such that 
\begin{equation}
E^s_{\bfl^{(j)}}<E, j=1,\cdots,k_E ,
\end{equation}
where 
\begin{equation*}
k_E\le A (\frac{E}{\Omega_0}),
\end{equation*}
$A$ is a finite geometrical constant, $\Omega_0$ appears in assumption (A2), and $E_{\bfl^{(j)}}^s$ is given in (\ref{SpecH0}). Let $|l|:=\max l_i.$ Then $|l^{(j)}|<\frac{E}{\Omega_0}$ for $j=1,\cdots ,k_E.$ Choose a contour $\gamma_{E}$ in the complex plane surrounding $\sigma (H_0(s))\cap [0,E),$ such that 
\begin{equation}
\label{dE}
d_E := \min_{s\in I} dist [\gamma_{E}, \sigma(H_0(s))]=\frac{1}{2}\min_{s\in I}(E^s_{\bfl^{(k_E+1)}}-E^s_{\bfl^{(k_E)}})>0.
\end{equation}
For each fixed time $s\in I$, we define the spectral projection of $H_1(s)$,
\begin{equation}
\label{Proj1}
P_E^{\theta,\epsilon}(s):= \frac{1}{2\pi i}\oint_{\gamma_{E}} dz (z-H_1(s))^{-1}.
\end{equation}
Let $P_E^0(s)$ be the orthogonal projection of $H_0(s)$ onto the subspace $\H_{E(s)}$ spanned by the eigenfunctions $\{ \phi_{\bfl^{(1)}},\cdots,\phi_{\bfl^{(k_E)}}\},$ and choose $\epsilon$ such that 
\begin{equation}
\label{AnalPertBd}
\epsilon c<\frac{d_E^2}{3(E+\Omega_0)},
\end{equation} 
where $c$ is a finite constant appearing in assumption (A3). It follows from analytic perturbation theory, with $\epsilon$ satisfying (\ref{AnalPertBd}), that 
\begin{equation}
Tr (P_E^{\theta,\epsilon}(s))=Tr (P_E^0(s))=k_E ,
\end{equation}
and 
\begin{equation}
\label{ProjDiff}
\| P_E^{\theta,\epsilon}(s)-P_E^0(s)\| <1 .
\end{equation}
We have the following lemma.

\bigskip

\noindent {\bf Lemma A.1} 

{\it Suppose assumptions (A2) and (A3) hold. Choose $\epsilon$ satisfying (\ref{AnalPertBd}), and fix $s\in I.$ Furthermore, suppose that $\psi^s\in Ran \; P_E^{\theta, \epsilon}(s).$ Then there exist finite constants $C>1$ and $\alpha>0$ (dependending on $\epsilon$) such that, for sufficiently small $\alpha,$
\begin{equation}
\label{Claim1}
\| e^{\alpha |x|^2}\psi^s \| \le C\| \psi^s\| .
\end{equation}
Furthermore, 
\begin{equation}
\| e^{\alpha |x|^2}U_1(s,s')\psi^{s'}\| \le C\| \psi^{s'}\|,\label{Claim2}
\end{equation}
for $\tau<\infty$ and $\alpha$ small enough.}

{\it Proof.} It follows from (\ref{ProjDiff}) that there exists $\phi^s\in \H_{E(s)},$ the subspace spanned by the eigenfunctions $\{ \phi_{\bfl^{(1)}},\cdots,\phi_{\bfl^{(k_E)}}\},$ such that 
\begin{equation}
\label{Psis}
\psi^s:= P_E^{\theta,\epsilon}\phi^s ,
\end{equation}
and hence 
\begin{equation}
\|\psi^s\| \le C \|\phi^s\| ,
\end{equation}
for some finite constant $C.$
Moreover, it follows from (\ref{Proj1}) and (\ref{Psis}) that
\begin{equation}
\label{ExpPsi}
e^{\alpha |x|^2}\psi^s = \oint_{\gamma_{E(s)}}  \frac{dz}{2\pi i} e^{\alpha |x|^2}(z-H_1(s))^{-1} e^{-\alpha |x|^2} e^{\alpha |x|^2} \phi^s .
\end{equation}
For $\alpha$ small enough, we know from (\ref{Hermite}) and (\ref{HermiteDecay}) that 
\begin{equation}
\label{ExpPhi}
\|e^{\alpha |x|^2}\phi^s\| \le C' \| \phi^s\| ,
\end{equation}
for some finite constant $C'.$ Moreover, for $z\in \gamma_{E},$ it follows from analytic perturbation theory, \cite{Ka1}, that
\begin{equation}
\label{Bd1}
\| e^{\alpha |x|^2}(z-H_1(s))^{-1}e^{-\alpha |x|^2}\| = \| (z-\overline{H}_1(s))^{-1})\| <\infty ,
\end{equation}
for $\alpha$ small enough (depending on $\epsilon$), where 
\begin{equation*}
\overline{H}_1(s):= H_1(s)+2\alpha d -4\alpha^2 |x|^2 +4\alpha x\cdot \nabla .
\end{equation*}
The claim (\ref{Claim1}) follows from (\ref{ExpPsi}), (\ref{ExpPhi}) and (\ref{Bd1}).
Now, 
\begin{equation*}
e^{\alpha |x|^2}U_1 (s,s')\psi^{s'} = \overline{U}_1(s,s')e^{\alpha |x|^2}\psi^{s'} ,
\end{equation*}
where $\overline{U}_1 = e^{\alpha|x|^2}U_1(s,s')e^{-\alpha |x|^2}$ is the propagator generated by $\overline{H}_1(s).$ By applying analytic perturbation theory, it follows that 
\begin{equation*}
\| \overline{U}_1(s,s')e^{\alpha |x|^2}\psi^{s'} \|\le e^{M(\alpha)\tau}\| e^{\alpha |x|^2}\psi^{s'}\| ,
\end{equation*}
where $M(\alpha)$ is a positive constant such that $M(\alpha)\rightarrow 0$ as $\alpha\rightarrow 0.$ Together with (\ref{Claim1}), this implies (\ref{Claim2}) for $\alpha$ small enough.

$\Box$
\bigskip

\bigskip
\noindent{\bf Proof of Proposition 3.1, Section 3}
\bigskip

{\it Proof of Proposition 3.1.} Fix $\theta,$ with $0<\Im\theta<\beta.$ By assumptions (B1) and (B3), there exists an open interval $I\subset \N(s,\theta)\cap {\mathbf R},$ with $\lambda_0(s)\in I.$ Choose $\xi\in C_0^\infty (I).$ Then
\begin{eqnarray}
F(s,t)&:=& \langle \psi_g^N(s), e^{-iH_g(s) t} \xi (H_g(s)) \psi_g^N(s)\rangle \nonumber \\
& = & \lim_{\epsilon\rightarrow 0}\int_I \frac{dz}{2\pi i} e^{-izt} \xi(z) \langle \psi_g^N(s), (R_g (s,z-i\epsilon)-\nonumber \\ &-&R_g(s,z+i\epsilon))\psi_g^N(s)\rangle .\label{F}
\end{eqnarray}  
Let 
\begin{equation}
f(\theta,s, t) := \frac{1}{2\pi i}\int_I dz e^{-izt}\xi (z) \langle \psi_g^N(s,\overline{\theta}), R_g (s,\theta; z) \psi_g^N(s,\theta)\rangle ,\label{f}
\end{equation}
where $\psi_g^N(s,\theta):= U(\theta)\psi_g^N(s).$
Then 
\begin{equation}
F(s,t)= f(\overline{\theta},s,t)-f(\theta,s,t) .\nonumber
\end{equation}
The resolvent in $\N(s,\theta)$ can be decomposed into a singular and regular part,
\begin{equation}
R_g(s,\theta;z)=\frac{P_g(s,\theta)}{z-\lambda_g(s)} + R_g^{analytic}(s,\theta;z),\label{Resolvent}
\end{equation}
where $R_g^{analytic}(s,\theta;z)$ is analytic in $z.$ Note that 
\begin{equation}
R_g^{analytic}(s,\theta;z)P_g(s,\theta) = P_g(s,\theta) R_g^{analytic}(s,\theta;z) = 0.\label{RegularRes}
\end{equation}
Using (\ref{RegularRes}), the contribution of the regular part to $f(\theta,s,t)$ defined in (\ref{f}) is 
\begin{equation}
\langle u_g^N(s,\overline{\theta}), \frac{1}{2\pi i}\int_I dz e^{-izt} \xi(z)R_g^{analytic} (s,\theta;z) u_g^N (s,\theta)\rangle,\nonumber
\end{equation}
where 
\begin{equation*}
u_g^N (s,\theta):= \frac{1}{\| P_g^N(s)\psi_0(s)\|}[P_g^N(s,\theta)-P_g(s,\theta)]\psi_0(s,\theta),
\end{equation*}
is of order $g^N.$ Since $\xi\in C_0^\infty (I),$ the last integral is bounded by $C_m t^{-m}$ for any $m\ge 0,$ and hence the contribution of the regular part is bounded by $g^{2N}C_m t^{-m}.$ The contribution of the singular part of the resolvent to $F(s,t)$ is 
\begin{equation}
\overline{a_g^N(s)} \frac{1}{2\pi i} \int_I e^{-izt} \xi(z) (z-\overline{\lambda_g(s)})^{-1} - a_g^N(s) \frac{1}{2\pi i} \int_I dz e^{-izt} \xi(z) (z-\lambda_g(s))^{-1} .\label{SingularContribution}
\end{equation}
Using the fact that $\xi=1$ in some open interval $I_0\ni \lambda_0,$ one may deform the path $I$ into two contours, $C_0$ and $C_1,$ in the lower complex half-plane, as shown in Figure 2. 

\bigskip
\begin{center}
\begin{picture}(0,0)%
\includegraphics{fig2.pstex}%
\end{picture}%
\setlength{\unitlength}{3108sp}%
\begingroup\makeatletter\ifx\SetFigFont\undefined%
\gdef\SetFigFont#1#2#3#4#5{%
  \reset@font\fontsize{#1}{#2pt}%
  \fontfamily{#3}\fontseries{#4}\fontshape{#5}%
  \selectfont}%
\fi\endgroup%
\begin{picture}(9339,3091)(1474,-4859)
\put(5986,-1951){\makebox(0,0)[lb]{\smash{{\SetFigFont{12}{14.4}{\rmdefault}{\mddefault}{\updefault}{\color[rgb]{0,0,0}$I_0$}%
}}}}
\put(6166,-3706){\makebox(0,0)[lb]{\smash{{\SetFigFont{12}{14.4}{\rmdefault}{\mddefault}{\updefault}{\color[rgb]{0,0,0}$C_1$}%
}}}}
\put(5356,-2671){\makebox(0,0)[lb]{\smash{{\SetFigFont{12}{14.4}{\rmdefault}{\mddefault}{\updefault}{\color[rgb]{0,0,0}$C_0$}%
}}}}
\put(6031,-2986){\makebox(0,0)[lb]{\smash{{\SetFigFont{12}{14.4}{\rmdefault}{\mddefault}{\updefault}{\color[rgb]{0,0,0}$\lambda_g$}%
}}}}
\end{picture}%

\end{center}
\bigskip

The term in (\ref{SingularContribution}) corresponding to the path $C_0$ picks the residue $a_g^N(s) e^{-i\lambda_g(s) t}.$ It follows from the identity
\begin{equation*}
P_g^N(s,\theta) P_g(s,\theta)P_g^N(s,\theta) = (P_g^N(s,\theta))^2 + [P_g^N(s,\theta)-P_g(s,\theta)][P_g(s,\theta)-1][P_g^N(s,\theta)-P_g(s,\theta)] ,
\end{equation*}
and from the fact that $$\| P_g^N(s,\theta)-P_g(s,\theta)\| = O(g^N),$$ 
that 
\begin{equation}
a_g^N(s) = 1 + O(g^{2N}).\label{ACoef}
\end{equation}
Using (\ref{ACoef}), one may write the remainder term in (\ref{SingularContribution}) due to the path $C_1$ as
\begin{align*}
\Im\lambda_g(s) \int_{C_1}\frac{dz}{\pi i} e^{-izt} \xi (z) (z-\overline{\lambda_g(s)})^{-1}(z-\lambda_g(s))^{-1} &+ O(g^{2N})\int_{C_1} dz e^{-izt} (z-\overline{\lambda_g(s)})^{-1} + \\ &+ O(g^{2N}) \int_{C_1} dz e^{-izt} (z-\lambda_g(s))^{-1} ,
\end{align*}
which is of order $O(g^{2N}).$ $\Box$

\bigskip



\end{document}